\documentclass[12pt]{article}
\usepackage{a4wide}
\usepackage{amssymb, amsmath}
\usepackage{hyperref}
\usepackage[normalem]{ulem}
\usepackage[dvips]{graphicx,color}
\usepackage[utf8]{inputenc}

\begin{document}
{\renewcommand{\thefootnote}{\fnsymbol{footnote}}
		
\begin{center}
{\LARGE Effective black-to-white hole bounces: \\
	The cost of surgery\\} 
\vspace{1.5em}

Suddhasattwa Brahma$^{1}$\footnote{e-mail address: {\tt suddhasattwa.brahma@gmail.com}} and Dong-han Yeom$^{1,2}$\footnote{e-mail address: {\tt innocent.yeom@gmail.com}}
\\
\vspace{0.5em}
$^1$ Asia Pacific Center for Theoretical Physics,\\
Pohang 37673, Korea\\
$^2$ Department of Physics, POSTECH,\\
Pohang 37673, Korea\\
\vspace{1.5em}
\end{center}
}
	
\setcounter{footnote}{0}

\newcommand{\bea}{\begin{eqnarray}}
\newcommand{\eea}{\end{eqnarray}}
\renewcommand{\d}{{\mathrm{d}}}
\renewcommand{\[}{\left[}
\renewcommand{\]}{\right]}
\renewcommand{\(}{\left(}
\renewcommand{\)}{\right)}
\newcommand{\nn}{\nonumber}

\def\be{\begin{equation}}
\def\ee{\end{equation}}

\begin{abstract}
We investigate possible geometries allowing transitions from a black hole to a white hole spacetime, by placing a space-like thin shell between them. Such proposals have been advanced recently to account for singularity-resolution in black-hole spacetimes. This space-like shell can be extended to be outside the event horizon and, thereby, reproduce some of the features of these proposals. On the other hand, if the space-like shell is confined fully within the horizon, then it results in a bounce near a space-like singularity inside the black hole. For both cases, the null energy condition is necessarily violated, at least effectively, due to introduction of quantum effects. If the shell, with  a non-trivial negative tension, extends beyond the event horizon, then one can see effects of quantum gravity modifications even outside the horizon as a cost of such a space-like surgery. Naturally, one needs to consider whether these types of manufactured spacetimes violates any known laws of nature, allowing for reasonable assumptions. After critically comparing our results with several models in the literature, we reiterate a new way to avoid such black-hole singularities without leaving a white-hole remnant via a quantum bounce.
\end{abstract}

\section{Introduction}
In classical general relativity (GR), gravitational collapse of matter, according to some initial data given on past null infinity, can lead to the formation of black holes \cite{Hawking:1969sw}. For matter sources satisfying reasonable null-energy conditions, it can be shown that one inevitably encounters a singularity due to the black hole. The existence of a singularity is a conceptual problem by itself, but when one considers the evaporation of a black hole \cite{Hawking:1974sw}, it generates more complicated problem -- the so-called `information loss paradox' \cite{Hawking:1976ra}.

There have been several candidate resolutions, for example, that Hawking radiation carries information \cite{Susskind:1993if}, that there forms a remnant or that the last outcome carries information \cite{Chen:2014jwq}, or perhaps that a bubble universe carries information \cite{Fischler:1990pk}. However, all these naive resolutions have some obstructions to being fully consistent \cite{Chen:2014jwq,Yeom:2008qw}. In order to overcome such inconsistencies, one may introduce a radical idea, e.g., a firewall \cite{Almheiri:2012rt} or the ER=EPR conjecture \cite{Maldacena:2013xja}, but these new proposals can lead to new problems \cite{Hwang:2012nn,Chen:2016nvj}.

Before resorting to such radical ideas, the conservative (and perhaps,  predominant) view is that the singularity occurs as a result of applying GR in regimes where a quantum theory of gravity should take over and thus such paradoxes can be ameliorated with the help of quantum effects. Indeed, putative quantum corrections are supposed to significantly alter the dynamics of matter collapse and therefore deviate from the standard picture of black hole formation, e.g., based on gravity \cite{Frolov:1988vj}, quantum field theory \cite{Stephens:1993an}, or loop quantum gravity \cite{Ashtekar:2005cj,InfLoss,LQGBH,Jibril}. However, there is no conclusive and unambiguous mechanism known to us yet as to how the regularization of the singularity might take place. Usually, a popular approach to avoid or resolve such a singularity is to introduce a bounce inside the black hole and effectively connect a black hole geometry to a white hole geometry, either using a de Sitter like vacuum core \cite{Frolov:1988vj} or due to (hypothetical) quantum gravitational effects \cite{Stephens:1993an,Ashtekar:2005cj,LQGBH}.

In the absence of a fundamental quantum gravity theory which allows us to evaluate the dynamical evolution of matter collapse scenarios in a systematic manner, there has been a lot of work recently, highlighted by several authors, as to achieve such a scenario effectively \cite{Haggard:2014rza,DAmbrosio:2018wgv,Christodoulou:2016vny,Bianchi:2018mml, Barcelo1,Barcelo2,Barcelo3}. The driving idea behind these proposals has been that quantum gravity would affect the matter collapse when they reach Planckian densities, thereby somehow overcoming the weight of the matter and leading to a bounce \cite{DAmbrosio:2018wgv}. To be more precise, the resulting geometry would be that of a black hole tunneling into a white hole quantum-mechanically \cite{Christodoulou:2016vny}, thereby leaving the white hole as a long-lived remnant \cite{Bianchi:2018mml}. The features common to all these models is that they \textit{assume} that quantum gravity effects, deep within the horizon, prevent the formation of an apparent trapping horizon into an event horizon through a quantum bounce. This requires that the (effective) null energy conditions are violated, at least for a finite region, due to quantum effects. Only in such a case would a bounce be available as is required by all such models. The other similarity between these proposals lie in the effective nature of their description which do not depend on the details of the fundamental quantum gravity theory, beyond its ability to cure the singularity with a bounce.

However, there are also crucial differences within these models. While in \cite{Haggard:2014rza} it is shown that the lifetime of such an event is very long, the authors of \cite{Barcelo1} are of the view that such a black to white hole transition must occur rather instantaneously, with a very short characteristic time scale. Not only are these important theoretical distinctions between these proposals\footnote{For instance, in the case of the bounce with a large time scale, it is argued that the renowned instabilities associated with the white hole leads to the disfavoring of the quantum-mechanical decay channel.}, but also lead to very different phenomenology based on them \cite{Barcelo2}.

In this work, we reasses the viability of these proposals through further consistency checks (note that there have already been some studies of instabilities, for instance in \cite{Barcelo2, ImprovedFW,CRPierre}, appearing in some of these individual proposals). Our goal is not to critique these individual proposals per se, but rather to examine the viability of such a \textit{smooth} transition between black and white holes. Our main technique in this quest would be the Israel junction conditions \cite{Israel:1966rt,Blau:1986cw} of spacelike hypersurfaces \cite{Balbinot:1990zz} (see also \cite{Chen:2017pkl}) which are used to `sew up' spacetimes via some hypersurfaces between them. Since all these models do not seem to destroy the local structure of spacetime in that they are still Riemannian, such a  strategy should be applicable and should indicate to us the extent of validity of Einstein's equations before quantum effects take over. Indeed, the bottom-up approach of all these models depend on metric-matching following quite a similar philosophy. 

As shall be demonstrated by us below, there are different ways in which such a `surgery' can be done to spacetimes describing matter collapse. Naturally, all of them have their merits and drawbacks. Interestingly, different gluing of spacetimes lead to different bounces, as described by these proposals, and the heuristic predictions also seem to be in agreement with our model, in each case. However, we also find that sometimes the severity of the cost for operating this surgery can result in unexpected distortions to classical spacetimes, in regions where we expect to have none. We shall conclude by highlighting a different proposal for a (non-singular) black hole model \cite{InfLoss,LQGBH,LQGSigchangeBH}, which results from a more top-down application of loop quantum gravity to black hole spacetimes, and can perhaps set us in the path of resolving some of these issues.

This paper is organized as follows. In Sec.~\ref{sec:jun}, we discuss the Israel junction equation and its generic solutions for space-like thin-shells. In Sec.~\ref{sec:bla}, we describe details of space-like shells, where one is extending to infinity and the other is limited to the event horizon. Based on these solutions, in Sec.~\ref{sec:com}, we critically revisit several proposals about effective bounces from black to white hole transition. Finally, in Sec.~\ref{sec:dis}, we summarize our results and comment in terms of quantum gravitational perspectives.

\section{\label{sec:jun}Junction equation of space-like thin-shells}
We first consider the possibility of having space-like thin shells as the junction between a black hole and a white hole spacetime \cite{Israel:1966rt}. To this end, we begin by reviewing the general junction conditions for such a scenario \cite{Balbinot:1990zz}.

\subsection{Equations of motion}
Let us restrict our attention to spherically symmetric spacetimes as we are interested in examining the Schwarzschild black hole. Given a matter field, we consider the space-like junction between two solutions
\begin{eqnarray}
ds_{\pm}^{2} = - f_{\pm}^{-1}(r) dr^{2} + f_{\pm}(r) dt^{2} + r^{2} d\Omega^{2},
\end{eqnarray}
where subscripts $+$ and $-$ denote the regions outside and inside the shell, respectively. At the junction surface, we impose the following space-like hypersurface
\begin{eqnarray}
ds_{\mathrm{shell}}^{2} = d\tau^{2} + r^{2}(\tau) d\Omega^{2}.
\end{eqnarray}

The tangent vectors (where, the coordinate convention is $[r, t, \theta, \varphi]$) can easily be evaluated as
\begin{eqnarray}
e^{\alpha}_{\;\tau} &=& \left( \dot{r}, \frac{\epsilon_{\pm}\sqrt{\dot{r}^{2} + f_{\pm}}}{f_{\pm}}, 0, 0 \right),\\
e^{\alpha}_{\;\theta} &=& \left( 0, 0, 1, 0 \right),\\
e^{\alpha}_{\;\varphi} &=& \left( 0, 0, 0, 1 \right),
\end{eqnarray}
where $\epsilon_{\pm} = \pm 1$, the significance of which shall be explained below. The dots above are with respect to the parameter $\tau$. Since the tangent vector on the induced metric is $u^{a} = (1, 0, 0)$, the corresponding tangent vector in the ambient spacetime can be written as
\begin{eqnarray}
u^{\alpha} = u^{a} e^{\alpha}_{\;a} = \left( \dot{r}, \frac{\epsilon_{\pm}\sqrt{\dot{r}^{2} + f_{\pm}}}{f_{\pm}}, 0, 0 \right),
\end{eqnarray}
and hence, if $\epsilon_{\pm} = +1$, then $t$ increases as $\tau$ increases, and vice versa (for $f_{\pm} > 0$). Later, this direction will be identified with the outward normal direction. From this $u^{\alpha}$, one can choose a normal vector $n_{\alpha}$ that satisfies $u^{\alpha}n_{\alpha} = 0$ and $n^{\alpha}n_{\alpha} = -1$, to get
\begin{eqnarray}
n_{\alpha} = \left( \frac{\epsilon_{\pm}\sqrt{\dot{r}^{2} + f_{\pm}}}{f_{\pm}}, -\dot{r}, 0, 0 \right).
\end{eqnarray}
Furthermore, one can then calculate the extrinsic curvature by the equations $K_{ab} \equiv n_{\mu;\nu} e^{\mu}_{\;a} e^{\nu}_{\;b}$. We obtain
\begin{eqnarray}
K^{a}_{\;b} = \left( {\begin{array}{ccc}
	\dot{\beta}_{\pm}/\dot{r} & 0 & 0 \\
	0 & \beta_{\pm}/r & 0 \\
	0 & 0 & \beta_{\pm}/r
	\end{array} } \right),
\end{eqnarray}
where $\beta_{\pm} \equiv \epsilon_{\pm}\sqrt{\dot{r}^{2} + f_{\pm}}$. 

The Israel junction conditions can be summarized as \cite{Balbinot:1990zz}
\begin{eqnarray}
[h_{ab}]&=&0, \\
\epsilon \left( \left[ K_{ab} \right] - h_{ab} \left[ K \right] \right) &=& -8\pi S_{ab},
\end{eqnarray}
where $\epsilon = +1$ for a time-like shell and $\epsilon = -1$ for a space-like shell. The notation adopted here is that, for any tensorial quantity $A$, $[ A ] = A|^{+} - A|^{-}$. Thus, it is clear from this that for a space-like thin shell smoothly joining two metrics, we not only require that the induced metric on this hypersurface matches on both sides but that the surface stress-energy tensor on the shell corresponds to the jump in the extrinsic curvature between the two sides. 

Note that the energy-momentum tensor for the induced metric $S_{ab}$ can be given by a perfect fluid of the form
\begin{eqnarray}
S_{ab} = \sigma u_{a} u_{b} + \lambda \left( h_{ab} + \epsilon u_{a}u_{b} \right),
\end{eqnarray}
where $u^{a} = (1, 0, 0)$ and $u_{a} = (- \epsilon, 0, 0)$ are the tangential vectors of the induced metric and
\begin{eqnarray}
S^{a}_{\;b} = \left( {\begin{array}{ccc}
	-\epsilon \sigma & 0 & 0 \\
	0 & \lambda & 0 \\
	0 & 0 & \lambda
	\end{array} } \right),
\end{eqnarray}
where $\sigma$ is the tension and $\lambda$ is the pressure. 

By plugging all components, the second junction equation and the energy conservation condition takes the form
\begin{eqnarray}
\epsilon_{-}\sqrt{\dot{r}^{2} + f_{-}} - \epsilon_{+}\sqrt{\dot{r}^{2} + f_{+}} &=& 4\pi r \sigma,\\
\dot{\sigma} &=& -2 \frac{\dot{r}}{r} \left( \sigma + \epsilon \lambda \right)
\end{eqnarray}
with $\epsilon_{\pm} = \pm 1$, where $+1$ if $r$ increases along the outward normal directions and $-1$ for the opposite case. These are the only two independent equations which one can derive. The other one (for the angular component) can be derived from these two and is given by
\begin{eqnarray}
	\frac{\epsilon_{-} \beta_{-} - \epsilon_{+} \beta_{+}}{r} + \frac{\epsilon_{-} \dot{\beta}_{-} - \epsilon_{+} \dot{\beta}_{+}}{\dot{r}} = 8\pi \lambda.
\end{eqnarray}

\subsection{Types of solutions}
For multiple matter fields satisfying the equations of state $\lambda_{i} = \epsilon w_{i} \sigma_{i}$, one gets
\begin{eqnarray}
\sigma(r) = \sum_{i} \frac{\sigma_{0i}}{r^{2(1+ w_{i})}}.
\end{eqnarray}
As an example, given a free scalar field, for both time-like and space-like shells, $\sigma > 0$ and $w = -1$ \cite{Chen:2015lbp}. 

The junction equation can be further simplified as \cite{Blau:1986cw}
\begin{eqnarray}\label{JuncEqn}
\dot{r}^{2} + V(r) &=& 0, \;\;\; \text{with}\\
V(r) &=& f_{+}(r) - \frac{\left( f_{-}(r) - f_{+}(r) - 16 \pi^{2} \sigma^{2}(r) r^{2} \right)^{2}}{64 \pi^{2} \sigma^{2}(r) r^{2}}.
\end{eqnarray}
From this, we can obtain the extrinsic curvature components as
\begin{eqnarray}
\beta_{\pm} \equiv \epsilon_{\pm} \sqrt{\dot{r}^{2} + f_{\pm}} = \frac{f_{-} - f_{+} \mp 16 \pi^{2} \sigma^{2} r^{2}}{8\pi\sigma r},
\end{eqnarray}
for the different signs of $\epsilon_{\pm}$.

\begin{figure}
	\begin{center}
		\includegraphics[scale=1]{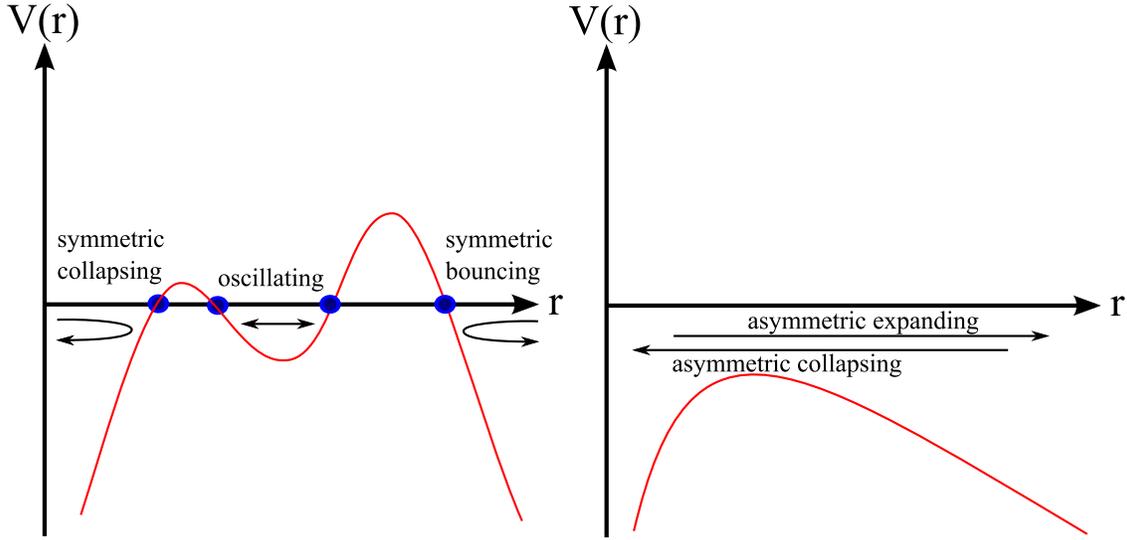}
		\caption{\label{fig:pots}Schematic figure of the effective potential $V(r)$ and types of solutions for symmetric collapsing, oscillating, and bouncing solutions (left) and asymmetric collapsing and expanding solutions (right).}
	\end{center}
\end{figure}

Since $V < 0$ in the allowed regions, there are different types of solutions depending on the location of zeros of $V(r)$ (Fig.~\ref{fig:pots}). If $V$ is negative for a region $0 \leq r \leq r_{1}$ or $r_{1} \leq r \leq \infty$, then the shell moves from $0$ to $r_{1}$ to $0$ or $\infty$ to $r_{1}$ to $\infty$. Hence, this solution is symmetric and can be classisified as either collapsing or bouncing. On the other hand, if $V$ is negative between two zeros, i.e., $V < 0$ for $r_{1} \leq r \leq r_{2}$, then an oscillatory solution  between $r_{1}$ and $r_{2}$ is possible. If $V$ is always negative, then a solution can cover from $0$ to $\infty$ (expanding) or $\infty$ to $0$ (collapsing), resulting in an asymmetric solution. Keeping this in mind, we can systematically classify all solutions into four different kinds \cite{Lee:2015rwa}:
\begin{itemize}
	\item[--] \textit{Symmetric bouncing solution}: starts from infinity, reaches to a minimum radius, and bounces back to infinity.
	\item[--] \textit{Symmetric collapsing solution}: starts from zero, reaches to a maximum radius, and bounces back to zero. 
	\item[--] \textit{Stationary or oscillating solution}: oscillates between two radius, where it is a stationary if two radius approach to the same value.
	\item[--] \textit{Asymmetric solution}: either collapsing from infinity to zero or expanding from zero to infinity.
\end{itemize}
The exact locations in the Penrose diagram should be carefully analyzed by comparing with the extrinsic curvatures since the signs of extrinsic curvatures will decide whether the shell locates the black hole side or the white hole side.

\section{\label{sec:bla}Black hole to white hole transition: Physical consequences?}
We shall consider two different geometries, considering a thin spacelike shell to patch up different spacetimes. In doing so, we shall illustrate the physical consequences for each of these models.

\subsection{A shell extending to infinity}
In order to consider a tunneling from a black hole to a white hole, we first choose
\begin{eqnarray}
f_{+}(r) = f_{-}(r) = \frac{2M}{r} -1,
\end{eqnarray}
and thereby set the convention
\begin{eqnarray}
\epsilon_{+} = 1, \;\;\; \epsilon_{-} = -1.
\end{eqnarray}
If there is to be a bounce which results in a transition from a black hole to a white hole, then at the point of the bounce there must be $\dot{r}=0$. When $\dot{r} = 0$, the left hand side of the junction equation \eqref{JuncEqn} is negative definite and hence $\sigma$ must be negative.

If we want to avoid such a condition, we can try to set $\sigma=0$. For this trivial condition on the stress-energy tensor, $\sigma=0$, the only solution is $\dot{r}^{2} = 1 - 2M/r$. In this case, $r \geq 2M$, for all $r$. This implies that this hypersurface is an Einstein-Rosen bridge, which is the same as $t = \mathrm{const.}$ hypersurface of outside the event horizon. Therefore, the Einstein-Rosen bridge is the only space-like hypersurface that we can paste without introducing a non-trivial thin-shell. 

\begin{figure}
	\begin{center}
		\includegraphics[scale=0.5]{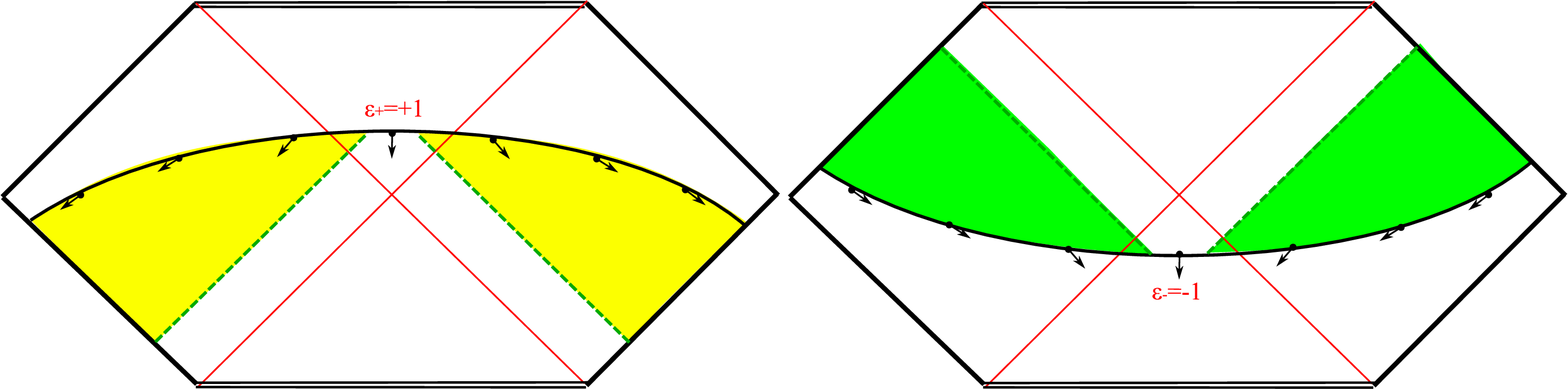}
		\caption{\label{fig:causal2}For the Haggard-Rovelli model, the left is for outside the shell and the right is for inside the shell. Note that this corresponds an asymmetric bouncing solution with $\epsilon_{+} = +1$ and $\epsilon_{-} = -1$ for all $r$. The green dashed lines corresponding an in-falling or out-going null shell.}
		\vspace{1cm}
		\includegraphics[scale=0.75]{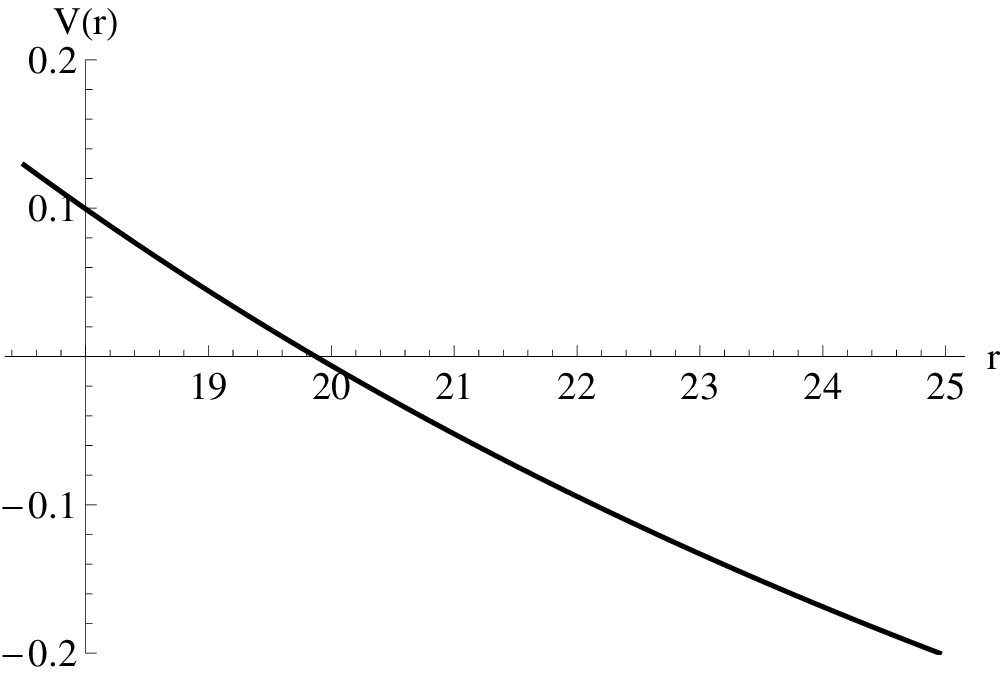}
		\includegraphics[scale=0.75]{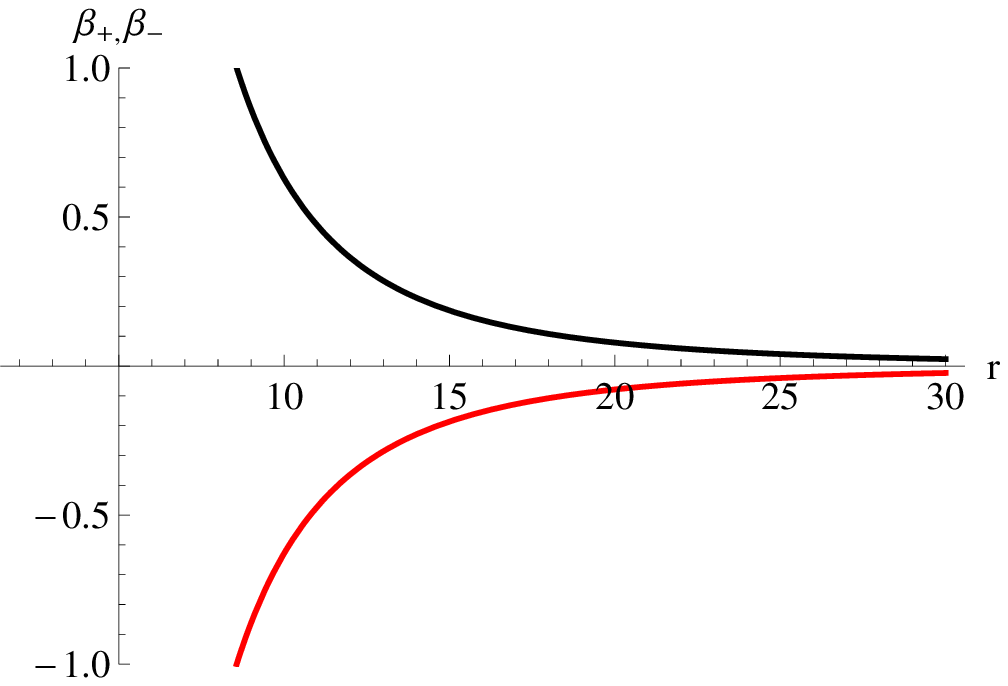}
		\caption{\label{fig:HR}Left: $V(r)$ for $M=10$, $\sigma_{0} = 100$, and $w_{1} = 1$. Right: $\beta_{+}$ (black) and $\beta_{-}$ (red).}
	\end{center}
\end{figure}

\begin{figure}
	\begin{center}
		\includegraphics[scale=0.5]{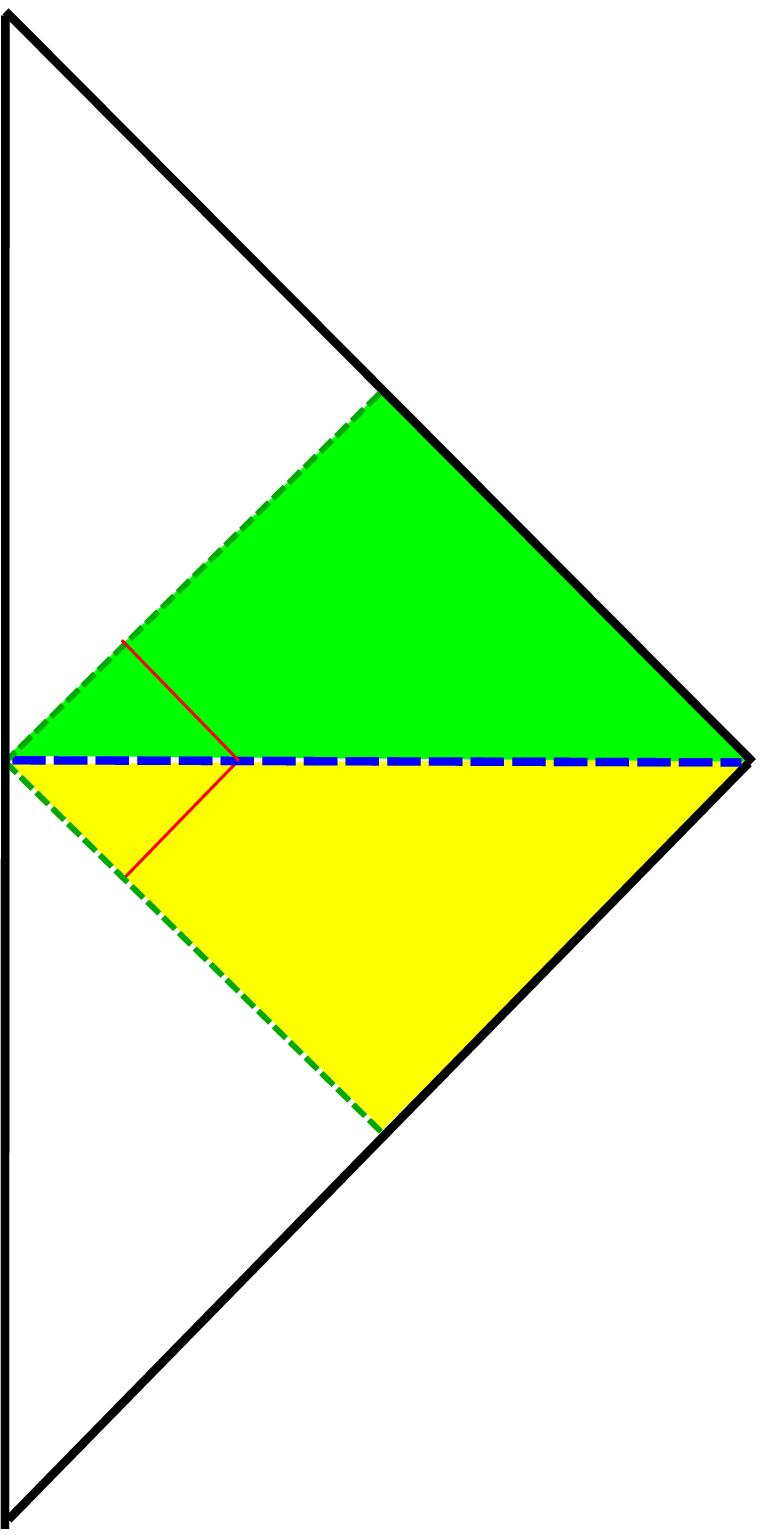}
		\caption{\label{fig:causal3}After pasting all hypersurfaces, there remains a hypersurface at infinity.}
	\end{center}
\end{figure}

Now, let us consider the non-trivial solution that covers $r < 2M$. If $V(r)$ allows a symmetric bouncing solution (starting from infinity, reaches a minimum radius inside the horizon, and moves to infinity again), then the causal structure looks like Fig.~\ref{fig:causal2}. This can easily be obtained, for example, by choosing $\sigma < 0$ and $w_{1} = 1$. Note that, if $\sigma < 0$ and $f_{-} = f_{+}$, then $\beta_{+} > 0$ and $\beta_{-} < 0$. Hence, for a $\sigma < 0$, we can always obtain a solution like in (Fig.~\ref{fig:HR}). After pasting all slices, the resulting Penrose diagram takes the form (Fig.~\ref{fig:causal3}).

If one expects to have a Riemannian geometry valid at all scales, then this seems the only way to glue the black and white hole spacetimes together, given the Israel junction conditions. One further assumption in our analysis has been that one uses a spacelike thin-shell to join these two regions (as opposed to, say, a null hypersurface). An immediate observation is that it is very easy to introduce such a model, for example, by requiring that $w > -1$ and we obtain $\sigma \sim 1/r^{2(1+w)}$. However, it also has some drawbacks as discussed in the next section (Sec.~\ref{sec:com}).

\subsection{A shell inside the horizon}
\begin{figure}
	\begin{center}
		\includegraphics[scale=0.5]{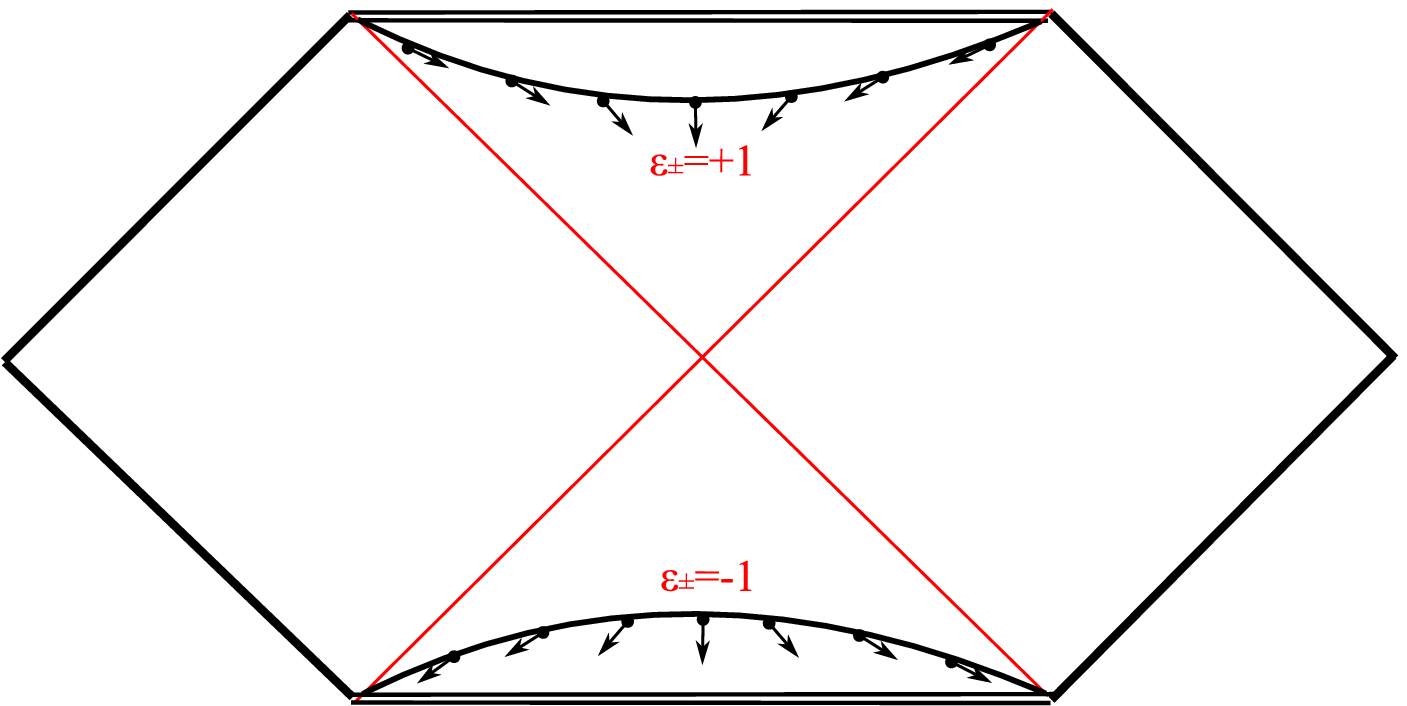}
		\includegraphics[scale=0.6]{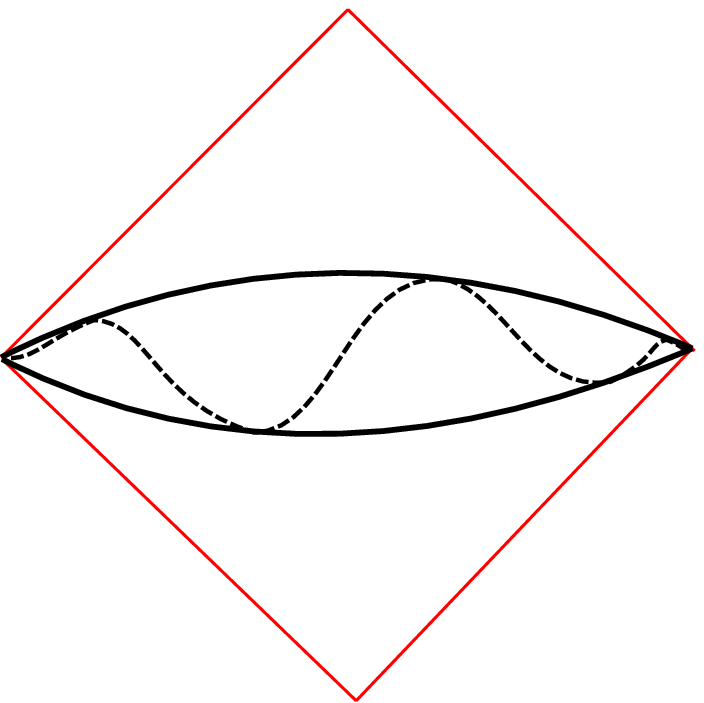}
		\caption{\label{fig:causal}Left: from top to bottom, the arrow directs the outward normal direction. Then the upper shell corresponds $\epsilon_{\pm} = +1$ while the lower shell corresponds $\epsilon_{\pm} = -1$. Right: even if the shell is in the oscillatory region, the solution still connects from a black hole to a white hole.}
	\end{center}
\end{figure}

We now go on to investigate another possible geometry classifying the transition from black to white holes. What if the junction joining the two regions is a complete Cauchy surface restricted fully inside the horizon? Then we do not need to have the difficulties mentioned above and are not required to pay the cost outside the horizon. Let us begin again with the equations
\begin{eqnarray}
f_{+}(r) = f_{-}(r) = \frac{2M}{r} -1,
\end{eqnarray}
with
\begin{eqnarray}
\epsilon_{+} = 1, \;\;\; \epsilon_{-} = -1,
\end{eqnarray}
valid only inside the horizon (Fig.~\ref{fig:causal}). Then, once again, $\sigma < 0$ is required. Now our purpose is to find a stationary or an oscillatory solution (Right of Fig.~\ref{fig:causal}) that is fully confined inside the horizon.

\begin{figure}
	\begin{center}
		\includegraphics[scale=0.5]{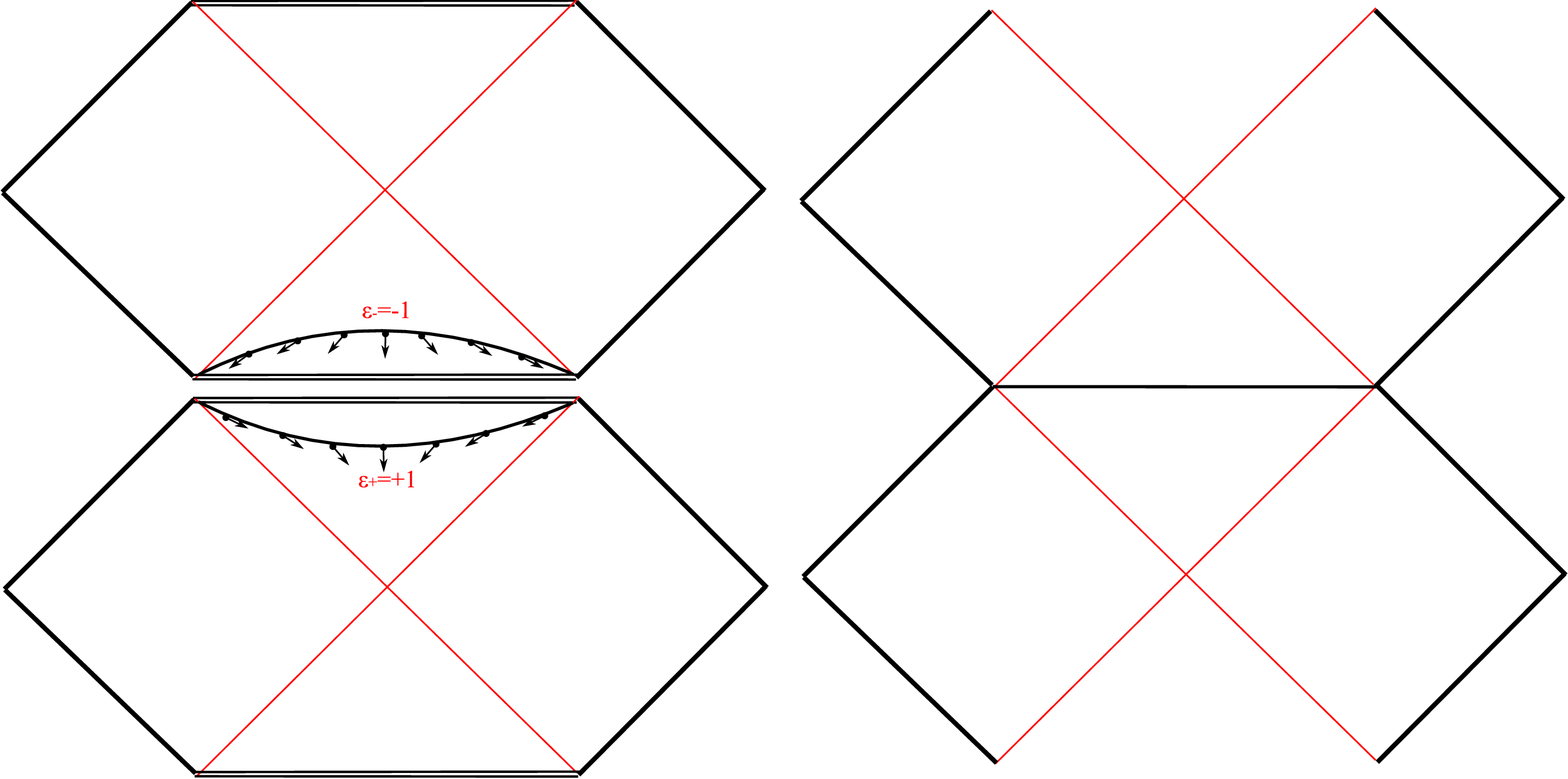}
		\caption{\label{fig:model2}Left: the inner shell (upper, $\epsilon_{-} = -1$) and the outer shell (lower, $\epsilon_{+} = +1$) will be pasted. Right: eventually, this causal structure describes a black hole to a white hole transition.}
		\includegraphics[scale=0.75]{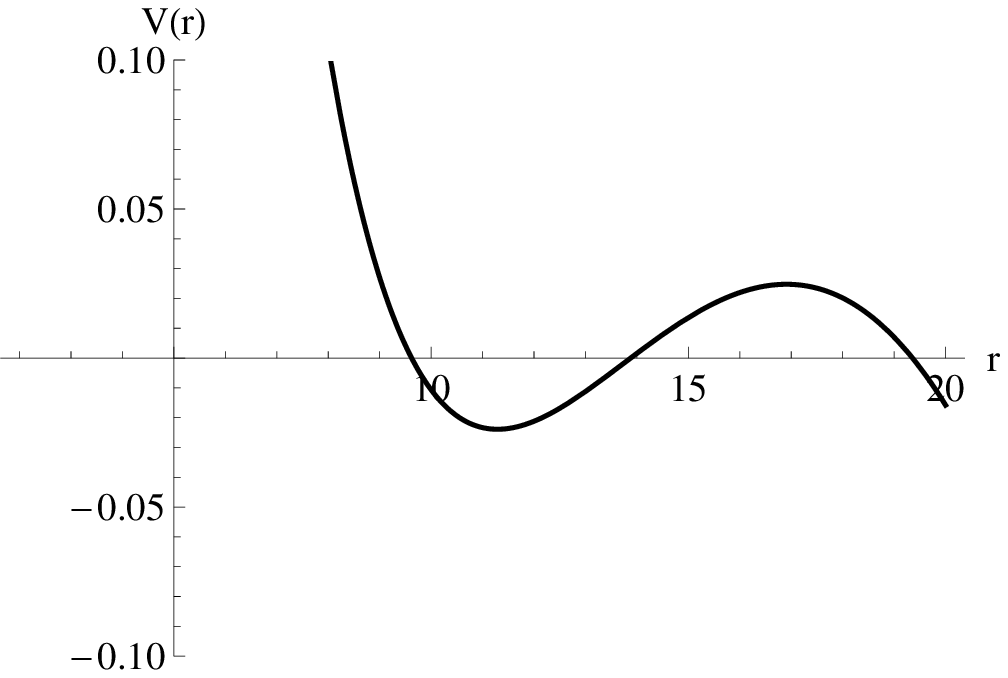}
		\caption{\label{fig:model2_pot}Left: $V(r)$ for $M=10$, $\sigma_{01} = 0.014$, $\sigma_{02} = -0.3$, $w_{2} = -0.5$. Right: extrinsic curvatures $\beta_{+}$ (black) and $\beta_{-}$ (red). Hence, $\epsilon_{+} = +1$ and $\epsilon_{-} = -1$.}
	\end{center}
\end{figure}

When $\dot{r} = 0$, since $(\sqrt{f_{-}} - \sqrt{f_{+}})/r$ is a negative and a concave function, in the limit $r \rightarrow 0$, $(\sqrt{f_{-}} - \sqrt{f_{+}})/r$ behaves $\sim r^{-3/2}$. Hence, for example, if we choose
\begin{eqnarray}
\sigma(r) = \sigma_{01} + \frac{\sigma_{02}}{r^{2(1 + w_{2})}}
\end{eqnarray}
with $w_{1} = -1$ and $w_{2} < -1/4$, then we can first obtain one zero $r_{1}$ near $r \sim 0$ such that  $V(r < r_{1}) > 0$. Therefore, $r_{1}$ can be a classical bouncing point. By tuning $\sigma_{01}$, one can find another zero $r_{2}$. Then $r_{1} < r < r_{2}$ becomes a classical oscillatory region. Once such a stable oscillatory region exists, the transition from a black hole to a white hole can easily be allowed (Fig.~\ref{fig:model2} and Fig.~\ref{fig:model2_pot}).

In this case, we find that the tangential equations of state are reasonable. However, in this case, the difficulty is with the kinetic term of the radial directions. The null energy condition should be violated and then behavior of the tension should be steeper than $r^{-3/2}$. The fine-tuning of parameters are sensitive up to the choice of $M$ and hence this may not be stable for an evaporating black hole.

\section{\label{sec:com}Comparison with other proposals}
We now go on to compare our results with those coming from two other approaches of black to white hole tunnelling which have been proposed recently.

\subsection{The Haggard-Rovelli model}
Recently, a model due to \cite{Haggard:2014rza} (HR) has suggested that one can get a black to white hole transition due to (loop) quantum gravity effects extending outside the horizon. In this case, a null shell of incoming matter is assumed to bounce well within the horizon due to quantum effects and then turns into a null shell of outgoing matter. In this case, it is assumed that the near-horizon geometry gets modified due to accrual of quantum effects originating outside the horizon. As a result, for a radial observer outside the horizon ($R > 2m$), the characteristic time scale for such a transition is very long due to relativistic time dilation. As a side effect, the usual semi-classical picture of black hole evaporation by Hawking radiation is consistent with such a black to white hole bounce. It is important to remember, however, that the proposed HR spacetime has to be mapped to (patches of) two different Kruskal manifolds. Although the resulting spacetime for the black to white hole geometry has the same form as in our model, Fig.~\ref{fig:causal2}, the way that they are constructed are rather different. In spite of this, we would like to examine the relative similarities and differences between these two approaches. 

\begin{figure}
	\begin{center}
		\includegraphics[scale=0.5]{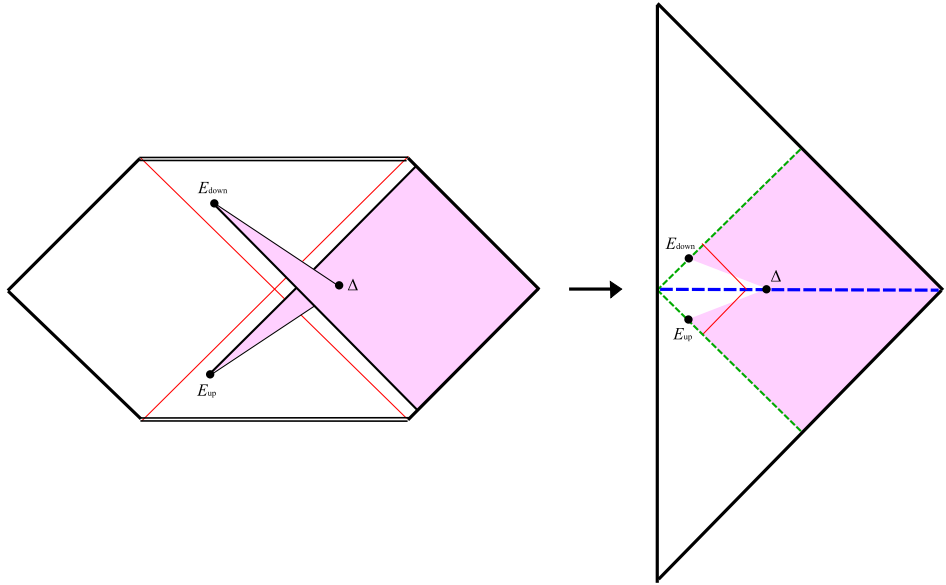}
		\caption{\label{fig:HR}Constructing the causal structure of the Haggard-Rovelli model.}
	\end{center}
\end{figure}

For our analysis, such a picture appears when $\sigma \neq 0$ (see Fig.~\ref{fig:HR}). Just as for our case, the HR model has a spacelike juntion between the black hole and its time symmetric (white hole) part. In this case, we find that there is a qualitative agreement between the conclusions between the HR model and our approach. We find that the solution for the stress-energy tensor which satisfies such a junction condition, we get $\sigma \sim 1/r^{2(1+w)}$, with $0 > w > -1$. This implies that, on this spacelike hypersurface, even outside the horizon, one must make allowances for such an effective negative tension. Note that such a negative energy component to the stress-energy tensor is expected well within the horizon so that one can avoid the singularity theorems and get a bouncing solution due to quantum gravity effects. However, it seems that there persists the remnant of such an effect well outside the horizon, if one considers the Israel junction conditions carefully. A nice feature of this is that at spatial infinity, $\sigma \rightarrow 0$ and hence it may be impossible to notice such a shell very far outside the horizon. However, as opposed to what has been conjectured in \cite{Haggard:2014rza}, we find that the effects of the shell are visible quite a bit outside the horizon. How physical such an effect can be a matter of debate.

In \cite{ImprovedFW}, it was pointed out that semiclassical stability requires a slightly assymetric bounce from a black hole to a white hole spacetime. This `improved fireworks' model, for our purposes, shares the same peculiarities of the original model as mentioned above. It is easy to generalize our approach to the improved fireworks scheme by identifying a spacelike hypersurface not precisely at $t=0$, but rather at some small positive $t$. In this case, there would be some consequences which would be different from the previous analysis. However, now the same junction conditions would be applied to this slightly shifted thin spacelike hypersurface. In conclusion, we would still require a negative tension on the hypersurface which gets smaller as one gets away from the horizon; however, there is still a remnant of this negative energy density outside the horizon due to quantum effects within the horizon. The model in \cite{ImprovedFW} does not claim otherwise.

\subsection{Exponential fading of black holes into white holes}
The common expectation for usual quantum bounces is that a UV theory would replace GR which would stop the matter collapse. In order to avoid the singularity theorems, one needs to either modify gravity or violate the energy conditions. However, even when one adds terms to the left hand side of Einstein's equation to reflect the modifications coming from the underlying quantum gravity theory, they can be transferred to the right hand side and used as an additional, `effective' stress-energy tensor contribution. In order to have a bounce, this effective stress-energy component must violate the null energy condition. This principle is also advocated in the works of Carballo-Rubio, Garay et. al. (see, for instance \cite{Barcelo3}).  

In this class of models, the authors advocate a minimum radius for the matter collapse emerging due to some (yet to be discovered) quantum theory of gravity. A smoothing procedure is then adopted such that there ends up being a negative stress-energy component. Due to the bounce of the incoming matter at the minimum radius, a `shock wave' is generated which then propagates outwards. This curvature shock wave can be made to have a finite radial extent by starting with a stellar structure with a finite radius. Once again, one finds that this shock wave would 
have some effect even outside the horizon although it originates at the higher curvature regime. This mechanism results in an exponential fading of black holes into white holes, for a co-moving observer. However, for our purposes, the important thing is that this resembles our model in that one finds something similar to a spatial hypersurface extending outside the horizon which remembers the effects of the UV theory, that resulted in the bounce. 

A further question arises whether, in this case, a quantum tunneling can realize such a smooth transition from a black hole to a white hole, whereby one gets an Oppenheimer-Snyder like collapse followed by a time-symmetric outward flow of matter after the bounce? By introducing a non-perturbative effect, such a transition seems possible in  principle. For example, in \cite{Sasaki:2014spa}, the authors considered a situation that a collapsing matter shell tunnels to a bouncing shell. Then, after tunneling, there remains neither a horizon nor a singularity. If one considers a thermal excitation of a shell \cite{Chen:2017suz}, then one can even obtain a tunneling channel which is independent of collapsing matter. However, this analysis shows that the tunneling probability is suppressed by the order of $e^{-M^{2}}$, where $M$ is the black hole mass \cite{Gregory:2013hja}. Unless one observes the entire wave function in the path-integral \cite{Hawking:2005kf}, these non-perturbative effects will be negligible, except for Planckian black holes. Therefore, it is not unreasonable to imagine that a collapsing shell tunnels into a bouncing shell when the black hole mass is around the Planck scale. However, this is not quite along the lines of the original proposal since the idea was to incorporate black hole of  classical size into this picture as well.

\section{\label{sec:dis}Discussion}
For a long time, it has been generally agreed upon that quantum gravity would solve the problem of black-hole singularity, which forms due to matter collapse. Recently, there has been a revival of the proposal that quantum effects would lead to the incoming shell of matter, bouncing off in the higher curvature regime, and then turn into an outgoing shell of matter. The quantum bounce is initiated due to some modification of gravity, which can be understood as an effective negative stress-energy component. There have been some inconsistencies, mainly in the form of instabilities, of these models which have already been pointed out in the existing literature \cite{Barcelo2, ImprovedFW, CRPierre}. Investigating the resulting geometry for the black to white hole transition, we further observe that 
\begin{itemize}
	\item[--] For the existing models in the market, we need to paste the past infinity (of the black hole) and the future infinity (of the white hole). Hence, we need a junction condition and there is no way to do this smoothly without introducing a  space-like shell. One can choose a condition such that $\sigma \rightarrow 0$ falls off at infinity, but not without extending a considerable effect outside the horizon. For instance, an effective ghost-like scalar field cannot be employed for this purpose. 
	\item[--] On the other hand, if the complete Cauchy hypersurface is restricted to be within the horizon, then a smooth transition without changing asymptotic infinity is possible, but the infinity of the black hole and the white hole cannot be located in the same universe. Also, we need to fine-tune the free parameters, resulting in such a choice being perhaps too delicate up to the change of mass.
\end{itemize}

Let us end with an example of quantum singularity-resolution which does not require a bounce or a black to white hole transition, as has been proposed by the models examined in this paper.  A novel way to avoid the singularity theorems would be if there is a new quantum spacetime which emerges in the higher curvature regime, in which the usual notions of (pseudo)-Riemannian geometry are not applicable any longer. In such a scenario, one need not violate the null energy condition even effectively to achieve singularity-resolution. Specifically, a black hole model proposed recently in loop quantum gravity \cite{InfLoss,LQGBH} has shown that a quantum Euclidean core emerges, in the deep quantum regime, due to holonomy modification functions from the theory. In this specific case, new geometric structures emerge and the usual notions of geodesic deviation (or, indeed of time-like geodesics) are not available any longer. However, curvature invariants remain bounded and one gets a non-singular quantum spacetime. On the other hand, there does not have to exist a negative stress-energy component, even effectively, to achieve singularity resolution in this case. Thus, there does not appear any need to have a geometry signaling a black hole to white hole transition for such a scenario.

The main question one would like to answer would be the fate of matter collapse and whether one can derive a consistent picture leading to the resolution of the black-hole singularity. The class of models examined in this paper seems to suggest that an incoming matter shell would bounce off in the deep quantum region, during the late stages of collapse, and eventually turn into an outgoing shell. Whether this process is symmetric, or what is the characteristic time for its completion as noticed by an observer, depends on the details of the construction. However, the main idea is that quantum gravity effects would lead to a bounce, due to an effective repulsive force, thus overcoming the formation of singularities, thereby demonstrating some form of `quantum hyperbolicity'. On the other hand, it might be such that one gets an Euclidean core inside the black hole which would imply that the details of the black hole evaporation process would depend on boundary conditions surround this region. In such models, one would require additional information on some future (Euclidean) boundary to traverse through the core, which is reminiscent of the final state condition as an additional requirement \cite{Final}. Another recent approach looks at avoiding singularities via a limiting curvature mechanism, resulting in a so-called `Russian doll' model of Schwarzschild spacetimes \cite{Mukhanov}. 

Thus, it far from clear what is the fate of dynamical matter collapse, in the presence of quantum gravity effects. In this work, we have tried to critically ascertain some of the features of bounce models, while reminding the reader that such mechanisms are not the only ones known to us for singularity resolution. One other avenue to be explored in the future would be to study the fate of matter collapse in a top-down approach of loop quantum gravity, where one finds the formation of Euclidean regions inside the black hole. Even for black to white hole bouncing models, another interesting option would be to study the construction of such geometries by pasting different spacetimes through thin null surfaces, instead of space-like ones.

\section*{Acknowledgment}
This work was supported in part by the Korean Ministry of Education, Science and Technology, Gyeongsangbuk-do and Pohang City for Independent Junior Research Groups at the Asia Pacific Center for Theoretical Physics.

\end{document}